\documentclass[twocolumn, a4paper]{article}

\usepackage{amsmath,amssymb,amsfonts}
\usepackage{algorithm}
\usepackage{algpseudocode}
\usepackage{algorithmicx}

\usepackage{graphicx}
\usepackage{textcomp}
\usepackage{xcolor}
\usepackage{color}
\usepackage{hyperref}
\usepackage{authblk}
\usepackage[left=1cm, right=1cm, top=2cm]{geometry}

\newcommand{\D}{\mathbb{D}}
\newcommand{\W}{\mathbb{W}}
\newcommand{\N}{\mathbb{N}}
\newcommand{\LHH}{L^{HH}}
\newcommand{\sHAP}{\textit{serving} HAP}
\newcommand{\sFSO}{\textit{serving} FSO}
\newcommand{\oHAP}{\textit{primary} HAP}
\newcommand{\bHAP}{\textit{backup} HAP}
\newcommand{\G}{$\mathbb{G}$}
\newcommand{\M}{$\mathbb{M}$}

\newcommand{\C}{\mathbb{C}}
\newcommand{\MHAP}{$\mathbb{M}^{HAP}$}

\newcommand{\AAnd}{\textbf{and }}
\newcommand{\AOr}{\textbf{or }}
\providecommand{\keywords}[1]
{
  \small	
  \textbf{\textit{Keywords---}} #1
}

\newcommand{\BER}[2][]{ \textrm{BER}_{#1#2}} 
\def \D	{\mathbb{D}}

\title{Survivable Free Space Optical Mesh Network using High-Altitude Platforms}

\author[1]{Dieu Linh Truong \thanks{linhtd@soict.hust.edu.vn}}
\author[1]{Xuan Vuong Dang}
\author[2]{The Ngoc Dang \thanks{ngocdt@ptit.edu.vn}}
\affil[1]{School of Information and Communication Technology, Hanoi University of Science and Technology, Vietnam}
\affil[2]{Department of Wireless Communications, Posts and Telecommunication Institute of Technology, Vietnam}

\date{}
\begin{document}
\maketitle

\begin{abstract}
Free space optical (FSO) communication refers to the information transmission technology based on the propagation of optical signals in space. FSO communication requires that the transmitter and receiver directly see each other. High-altitude platforms (HAPs) have been proposed for carrying FSO transceivers in the stratosphere. A multihop HAP network with FSO links can relay traffic between ground FSO nodes. In this study, we propose an end-to-end switching model for forwarding traffic between massive pairs of ground FSO nodes over a HAP network. A protection mechanism is employed for improving the communication survivability in the presence of clouds, which may break the line of sight (LoS) between HAPs and ground nodes. We  propose an algorithm for designing the topology of the survivable HAP network, given a set of ground FSO nodes. The results demonstrate that, even though networks with survivable capacity use more resources, they are not necessary much more expensive than those without survivability in terms of equipment, i.e., HAPs and FSO devices, and in terms of wavelength resource utilization. 
\end{abstract}

\section{Introduction}
\label{sec:intro}

Free space optics (FSO) communication technology uses a laser beam to transmit information through air or vacuum between a pair of transceivers at high speed. Commercial FSO transceivers available in the market usually operate at 1.25 Gbps, 2.5 Gbps or even higher speeds. However, FSO communication requires line of sight (LoS) between transceivers, which is sometimes difficult to satisfy, mostly at long distances on the ground. In addition, terrestrial FSO links are sensitive to weather conditions such as rain, fog, or air turbulence leading to link failures.

Some studies have addressed terrestrial FSO link failures by building terrestrial FSO networks with spare capacity, for example, the works in \cite{truong2017, robustFSO-2017, robustFSO-2019}. These works propose network dimensioning solutions, taking into account single link failures, multiple link failures, and partial link failures.

\begin{figure}
\includegraphics[width=0.5\textwidth]{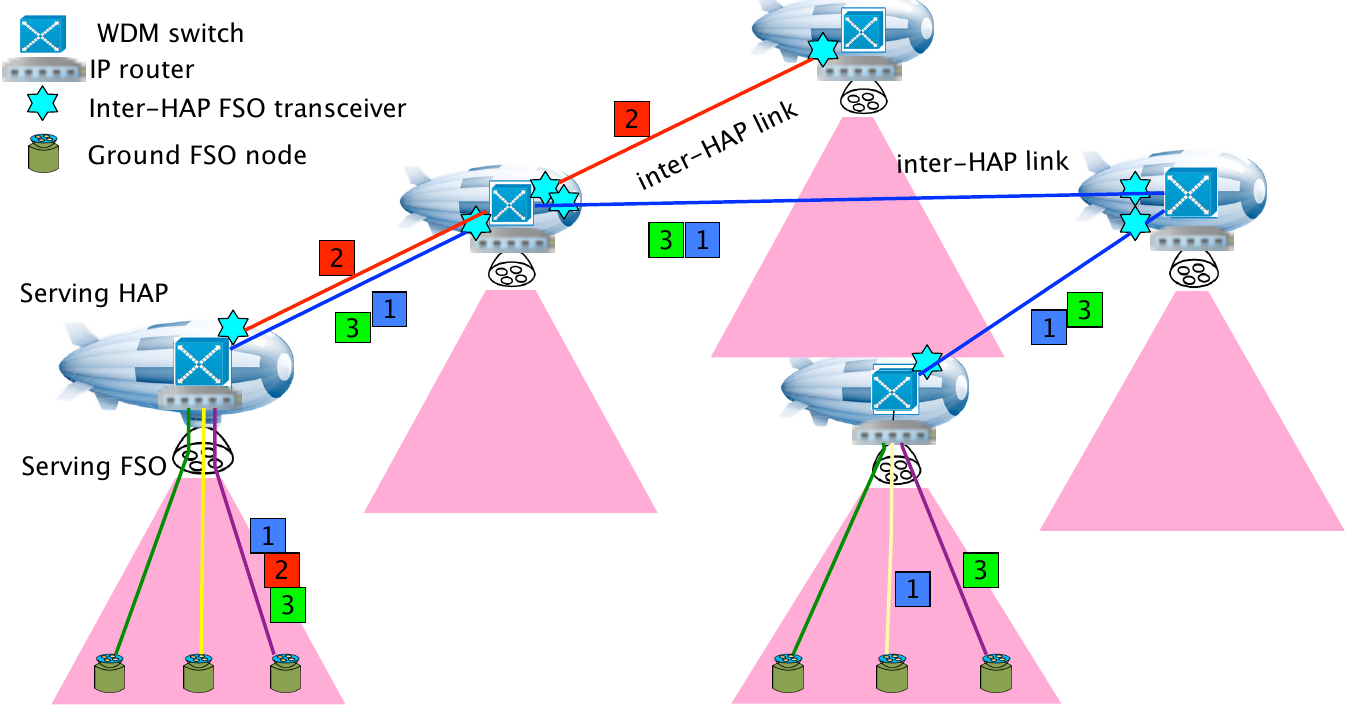}
\caption{Data communication in a FSO network using HAPs}
\label{fig:commHAP}
\end{figure}

Recent research has proposed the use of high-altitude platforms (HAP) for wireless communications, including FSO communication at different scales \cite{HAP-for-Wireless2001, Knapek2007, Mobile-HAP-2008, HAP-FSO-Sharma-2016}. HAPs are high-flying aircraft or airships, which operate at altitudes of 17 to 24 km in the stratosphere. HAPs are used to avoid physical obstacles and weather impact present on the ground. In stratosphere, HAPs suffers the least wind impact thus less power is required to stabilize the  platform. HAPs carry FSO transceivers that play the role of intermediate nodes forwarding traffic in the stratosphere or between the stratosphere and the ground. Figure \ref{fig:commHAP} illustrates the communication between the ground FSO transceivers (also called ground FSO nodes) using HAPs. The traffic between a pair of ground FSO transceivers follows a path comprising three parts: i) \emph{uplink}: from the source-ground FSO node up to an FSO transceiver on a HAP, ii) \emph{inter-HAP links}: between FSO transceivers on different HAPs, iii) \emph{downlink}: from the last HAP down to the destination-ground FSO transceiver. A HAP can communicate with several ground FSO transceivers. Although inter-HAP links are not effected by weather conditions, uplink and downlink suffer interference from clouds floating at an altitude of approximately 10km.

Several research projects on designing HAPs to deploy broadband networks such as Halo, HeliNet, CAPANINA, SkyNet, Loon project etc., have been conducted worldwide. Some reviews can be found in \cite{HAP-review-2007, HAP-survey-2005, Optical-HAP-2010, LoonProject, FB2017}. A number of technical reports have been produced from these projects on general network architecture requirements \cite{CapaninaD13} as well as physical interface specifications \cite{CapaninaD08}. In general, HAP is not only costly but also limited in capacity in terms of payload weight and continuous flying time before landing to recharge. Therefore, the utilization of HAPs should be well planned.

Other studies focus on the utilization of HAPs. In \cite{Mobile-HAP-2008}, each HAP carries a mobile based station that serves a cluster of mobile ground nodes. The work proposed a solution to divide ground nodes into clusters but the connectivity between HAPs has been omitted from the scope of this research. In \cite{Moho2006}, the authors proposed an optical transport network based on mesh HAP systems with inter-HAP links using FSO technology, whereas radio frequency (RF) was used between HAPs and the ground. All-optical data transmission with wavelength routing was used between HAPs. Nodes on HAPs are capable of MUX/DEMUX data received from the ground using a time devision multiplexing (TDM) technique. The authors agreed that the number of required wavelengths on the inter-HAP links and the wavelength routing scheme depend on the actual physical HAP topology. However, they only evaluated the maximum number of required wavelengths on  two very particular network topologies: bus and full mesh with five HAPs where each HAP only serves one ground station. The possibility that a HAP serves more than one ground station has not been clearly discussed. 

European studies on the availability of optical links between HAPs and ground in \cite{CloudHAP2007} show that single optical link availability varies from under 20\% during winter in Northern Europe to over 70\% during summer in Southern Europe. Research has focused on improving the availability of single optical downlink to ground stations by using a ground-station diversity scheme, i.e., ground stations at geographically separate locations. Availability increases with the number of available stations. The scheme requires a HAP network to carry traffic toward available ground links. Four ground stations in Southern Europe can achieved 99\% availability. Nevertheless, ground-station diversity is applicable only if the location of the transmission destination is not important. 

In \cite{Knapek2007}, data from Earth observation satellites were relayed to some HAPs at 20 km altitude through FSO links before sending them down to ground stations via microwave point-to-point links. Since there are no clouds at an altitude of 20 km, the availability of the FSO links is almost 100\%. In cloud-free conditions, FSO links can also be used in place of microwave links for a higher data rate.  

Research in \cite{HAP-FSO-Sharma-2016} proposed multi-HAP networks for carrying data between ground FSO nodes and optical slant links for transmitting data up to and down from HAPs when primary connections fail owing to air turbulence. This research focused only on analyzing the bit error rate (BER) and the availability of a multihop FSO link. The choice of HAP locations, slant links, connectivity between HAPs and routing from ground nodes to ground nodes were left outside of the study. 

Other studies either focused on analyzing the BER of multihop HAP-based FSO connections \cite{ngoc2019} or HAP network dimensioning without survivability against LoS loss \cite{TruongNICS2019}. 

In summary, to the best of our knowledge, there is no work on the dimensioning of HAP-based FSO networks that are survivable against optical HAP-ground link failures. We address this problem in this study. The network is composed of HAPs carrying FSO transceivers in the stratosphere for forwarding optical data between ground FSO nodes, where all links including uplinks and downlinks use FSO technology. Similar to \cite{HAP-FSO-Sharma-2016} we use FSO slant links for survivability. We do not elect the RF technology for slant links because RF links offer much lower data rate than optical links leading to connection quality down-grading when failures occur. Unlike previous studies, we focus on the dimensioning of the HAP-ground network.
There are two main contributions in this paper: i) proposing an end-to-end data switching mechanism between massive pairs of ground FSO nodes through a HAP network with a 1+1 protection mechanism against uplink and downlink failures, ii) a dimensioning solution for the survivable HAP-ground network with the minimum cost of equipment including HAPs and FSO transceivers costs. 

The remainder of this paper is organized as follows. Section \ref{sec:communication} presents the switching mechanism from end-to-end and the protection mechanism against clouds. Section \ref{sec:problem} describes the network dimensioning problem. Section \ref{sec:topo} presents the proposed dimensioning algorithm. Section \ref{sec:simulation} presents the simulation results and discussions. Finally, Section \ref{sec:conclusion} concludes the paper.

\section{Switching and Protection mechanisms}
\label{sec:communication}

\subsection{WDM communication between HAP and ground}
Each HAP carries one down-facing FSO device to communicate with a cluster of ground FSO transceivers. For the ground FSO transceivers, the HAP is called the \sHAP~and the FSO device on the HAP is called the \sFSO~transceiver. On the same HAP, there may be other FSO transceivers for communicating with other HAPs. These FSO transceivers are called \textit{inter-HAP} FSO transceivers. The \sFSO~transceiver and \textit{inter-HAP} FSO transceivers on the same HAP should be wired together back-to-back  using optical fibers (see Figure \ref{fig:commHAP}).

We divide the ground FSO transceivers into clusters, each having a \sHAP. Each ground FSO transceiver communicates with its \sFSO~bidirectionally by using two fixed wavelengths for the up and down directions. The wavelengths in the two directions are managed separately because each FSO transceiver has  a light source and a light receiver for sending and receiving data independently. The ground FSO devices of the same cluster share access to the common \sFSO~transceiver using wavelength devision multiplexing (WDM) technology. Figure \ref{fig:commHAP} illustrates this multiple-access method, where each color represents a wavelength. For simplification, we show only the wavelengths for the up or down direction. Depending on wavelength density of the WDM technique to be used, 32, 64 or more ground FSO transceivers can be served by a single FSO transceiver on HAP.

On \textit{inter-HAP links}, WDM technique will also be used. Between HAPs data are forwarded on a wavelength-switching basis without wavelength conversion. Data sent from one HAP to the other may travel through multiple HAPs using a continuous wavelength, the so-called lightpath. When the volume of traffic between a pair of HAPs is larger than the capacity of a wavelength, multiple lightpaths may be necessary. These lightpaths do not need to follow the same route.

\subsection{Data switching on HAP}

A ground FSO transceiver may have data to send to different remote ground FSO transceivers. This section explains how the data are switched from end to end. We propose to employ one of two models: IP over WDM and SONET over WDM.

In \textit{IP over WDM} model, the ground FSO nodes in the same cluster must be assigned IP addresses of the same network. An IP router is installed on each HAP to groom traffic received from the \sFSO~transceiver based on destination clusters. Then, the groomed traffic follows different lightpaths between HAPs to reach destination clusters. The end-to-end data are switched as follows:
\begin{itemize}
\item At a source ground FSO node, data are encapsulated in IP packets, of which the destination IP addresses are those of the destination ground FSO nodes. 
\item The packets follow an uplink and arrive at the \sFSO~of the source. Here, the IP packets pass through an IP router that directs the packets to different ports dedicated to different destination cluster networks. 
If the amount of data addressed to a cluster network is larger than the capacity of a wavelength then multiple ports should be used for that cluster.\\
The flow of IP packets going out from each router port is regenerated in optical domain using a wavelength and then forwarded to a WDM switch to take a lightpath to get to the \sHAP~of the destination cluster. 
\item At the \sHAP~of the destination cluster, IP packets inside each lightpath are DEMUX by the IP router of the HAP according to  their destination IP addresses. IP packets heading to the same destination ground FSO node are regenerated in optical domain, using the wavelength assigned to the destination ground FSO node, and transmitted to the ground.
\end{itemize}

\begin{figure}
\includegraphics[width=0.5\textwidth]{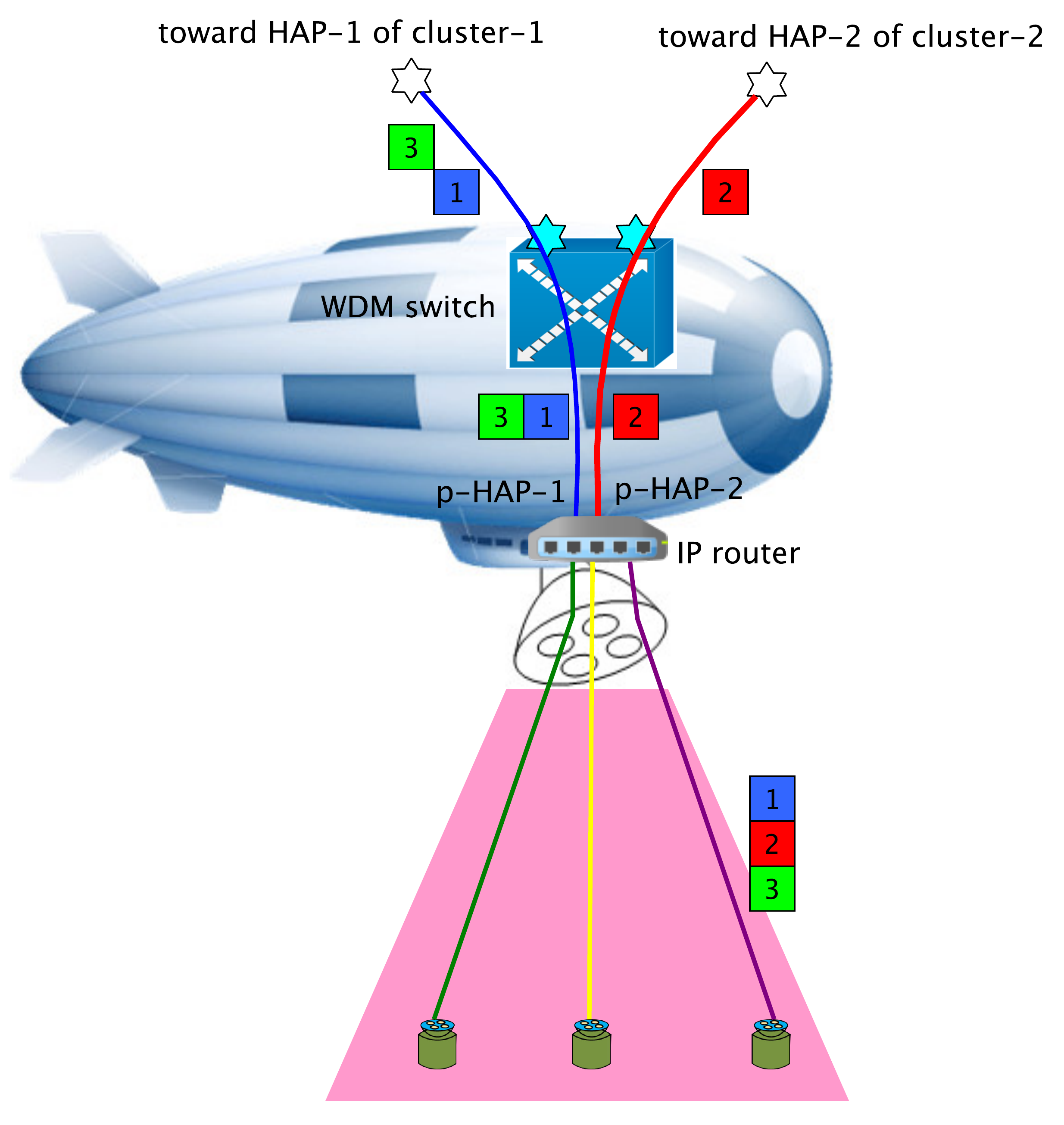}
\caption{Data switching on HAP. }
\label{fig:switchingHAP}
\end{figure}

In this switching mechanism, the set of lightpaths between HAPs should be pre-configured beforehand. These lightpaths can be considered as physical links between IP routers. Figure \ref{fig:switchingHAP} illustrates data switching on a HAP. Port p-HAP-1 is dedicated to receiving traffic for HAP-1 to get to network cluster-1. Port p-HAP-2 receives traffic for HAP-2 to get to network cluster-2. Table \ref{tab:routing} shows the IP routing table of the router in Figure \ref{fig:switchingHAP}.  
 \begin{table}[htb]
 \begin{tabular} {l|l}
 \hline
 Network	&	Interface \\
 \hline
 Network address of cluster-1	& p-HAP-1 \\
Network address of cluster-2	& p-HAP-2 \\
 \ldots 	& \ldots\\
 \hline
 \end{tabular}
 \caption{Routing table of the IP router in Figure \ref{fig:switchingHAP}}
 \label{tab:routing}
 \end{table}
 
In \textit{SONET over WDM} model, SONET switches replace IP routers. While IP routers forward data by packet switching basis, SONET switches work on TDM circuit switching basis. Therefore, a circuit should be established for each pair of ground FSO nodes by taking a fixed time slot inside the lightpath between the associated source and destination \sHAP s. The data rate of the time slot corresponds to the volume of requested traffic between the two ground FSOs.

Once one of the above models is employed, the remaining question is setting up lightpaths between HAPs and configuring the IP routers/SONET switches for grooming traffic to lightpaths. In Section \ref{sec:topo}, we propose algorithms for building the HAP topology and the lightpaths between HAPs, i.e., their paths and wavelengths. Subsequently, IP routers/SONET switches can be easily configured.

\subsection{Protection model}
\label{sec:protection}
On cloudy days, clouds may cut the LoS between a ground FSO node and its \sFSO~device. Clouds generally float  at an altitude of 10 km. To maintain continuous communication in the network, we propose the following protection mechanism against clouds. 

When the LoS between ground FSO nodes and their \sFSO~are cut, these ground nodes should forward all their traffic to a nearby alternative HAP, for which the LoS is still clear. Let \oHAP~be the \sHAP~where data travel in case of no failure whereas the alternative HAP is \bHAP. The \bHAP, after receiving data from a ground FSO node on a wavelength, will simply forward that wavelength to the \oHAP~without any processing. At the \oHAP, traffic is handled exactly as it is received directly from the ground, i.e. passing through the IP router and continuing the original lightpaths to the destinations. To receive the backup traffic  from the ground, each \bHAP~should have a dedicated FSO device that looks at its backup zone. This FSO device is called \textit{backup-serving FSO}.

\begin{figure}
\includegraphics[width=0.5 \textwidth]{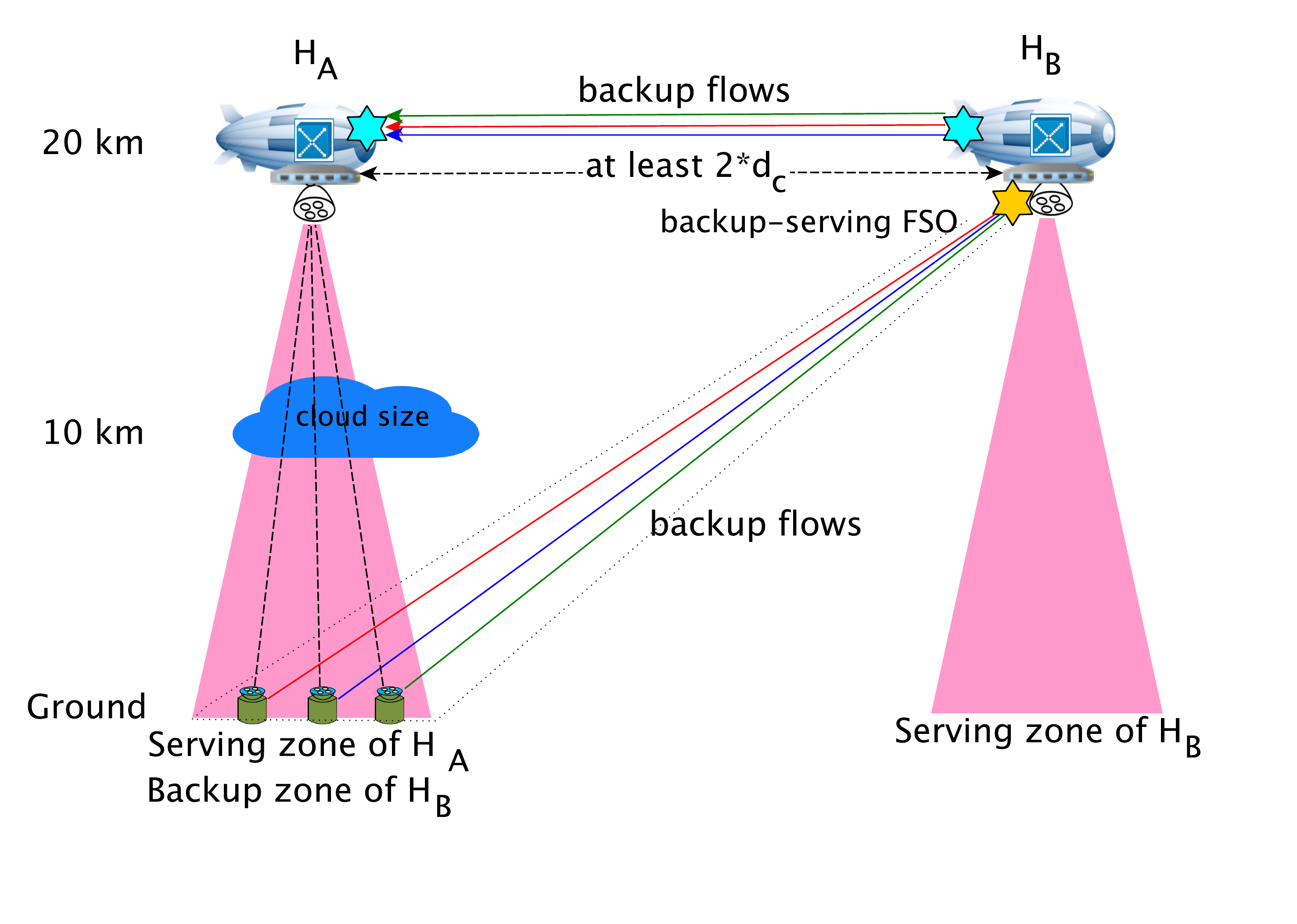}
\caption{Protection model: when LoS between $H_A$ and its serving zone is cut, every ground node in serving zone of $H_A$ should forward its traffic to $H_B$. \textit{Backup} HAP~should be at a distance of at least 2 times the size of cloud from its \oHAP.}
\label{fig:backupsol}
\end{figure}

Figure \ref{fig:backupsol} illustrates the protection model where zone A is the serving zone of HAP $H_A$ and HAP $H_B$ backs up for $H_A$. Every ground node in the serving zone of $H_A$ should forward its traffic to $H_B$ whenever its LoS to $H_A$ is cut. $H_B$ has a \textit{backup-serving FSO}~device for covering zone A. 

To implement this model, each ground node should have two light sources and two light detectors oriented toward $H_A$ and $H_B$. Such an FSO device has been mentioned in \cite{HAP-FSO-Sharma-2016}. We suggest that the ground nodes transmit two copies of data to both $H_A$ and $H_B$, and $H_B$ always forwards the copies it receives to $H_A$. $H_A$ chooses to receive the traffic directly from ground nodes or from $H_B$ depending on the availability of LoS between $H_A$ and its serving zone. This is a 1+1 backup mechanism. 

For downstream flows, the backup flows travel along the same paths with the upstream flows but in the reverse direction. When $H_A$ has to forward data to its ground FSO nodes, it transmits a copy to $H_B$ to relay the copy to the ground FSO nodes. It is the ground node that decides whether to receive from $H_A$ or $H_B$.

A \bHAP~$H_B$ must be sufficiently far from its \oHAP~$H_A$ to avoid that a cloud cuts both links to 
\oHAP~and \bHAP. Based on the altitude of HAPs and clouds, we deduce that the minimum distance between $H_A$ and $H_B$ should be twice the maximum size of clouds. However, a \bHAP~should not be too far from its backup ground FSOs because greater distance causes greater attenuation and air turbulence impact leading to high BER. Therefore, the distance $d(H_A, H_B)$ between $H_A$ and $H_B$ is restricted by:
\begin{equation}
	2 \times d_c  < d(H_A, H_B)< \LHH   \label{eq:d}\\
\end{equation}
where $d_c$ is the maximum cloud size, $\LHH$ is the maximum length of inter-HAP links with BER under the requirement threshold of the current Forward Error Correction techniques. 

It is worth noting that if $H_B$ backs up $H_A$, then topologically, $H_A$ could also backup $H_B$. 


Although the links from a ground node to its \oHAP~and \bHAP~are not influenced simultaneously by the same cloud, they can still be interrupted by two independent clouds. Let the probability that a link between a ground node and a HAP is cut by clouds be $p_{cut}$, then the probability that both links from a ground node to its \oHAP~and \bHAP~are cut by two independent clouds is $p_{cut}^2$. Consequently, the probability that one of the two links is available for transmission is $1-p_{cut}^2$. 

The following two cases illustrate the availability offered by the proposed protection mechanism. In the Southern of Europe, the availability of a single link  varies between 50\% -- 85\%~over a year \cite{CloudHAP2007}. Therefore, the proposed protection mechanism allows to increase the joint availability of the links to \oHAP~and \bHAP~to 75\% -- 97,75\%, which is relatively good. However, in the Northern of Europe, the mean availability of a single link is lower than 40\% then joint availability of the two links is still under 64\%. More than one slant link may need be considered for improving the joint availability with trade off in cost of backup HAPs and backup FSO transceivers. However, in this research, we focus on single backup slant link only.

\section{Survivable ground-HAP network dimensioning problem statement}
\label{sec:problem}

In this section, we propose a network dimensioning solution that identifies the topology of the HAP network, as well as the routing between ground FSO nodes over that network. The dimensioning solution aims to minimize the investment cost in terms of HAPs and FSO devices while satisfying the constraints for all traffic demands to be backed up against clouds. 
In order to model the impacts of air turbulence, we use Gamma-Gamma distribution model proposed in \cite{ngoc2019} for both ground-HAP and inter-HAP channels. The impacts are represented as BERs of ground-HAP and inter-HAP links. Then, while dimensioning the ground-HAP network, we requires that all lightpaths between HAPs must have end-to-end BER under a threshold $\delta$ such that bit errors could be corrected by current FEC techniques.

This dimensioning problem is stated as follows.\\
Given: 
\begin{itemize}
\item A set of ground FSO devices.
\item \M: the matrix of traffic demands between ground FSO devices.
\item Some other parameters as listed in Table \ref{tab:param}.
\end{itemize}
Output: A survivable topology using the proposed protection model that satisfies: 
\begin{itemize}
\item All traffic demands are routed, i.e, there exist a serving HAP for each ground FSO node and an inter-HAP lightpath for carrying each demand.
\item All traffic demands are backed up.
\item From HAPs to HAPs, data are routed over continuous lightpaths with end-to-end BER under a threshold $\delta$.
\item The total investment cost is minimized
 \begin{equation}
 	\min (n_{HAP} * c_{HAP} + n_{FSO} * c_{FSO})
 \end{equation}
\end{itemize}
where $n_{HAP}$ and $n_{FSO}$ are the number of HAPs and the total number of \sFSO~installed on HAPs, respectively. $c_{HAP}$ and $c_{FSO}$  are the investment costs for a HAP and an FSO on HAP, respectively.

In this research, we assume that the apertures of \sFSO s are identical, thus their coverages on the ground are fixed by $\D$.

\begin{table}
\begin{tabular}{l|l|l}
\hline
Param. 	& Description	&Sim. value \\
\hline
$\N$ 		& Set of ground FSO transceivers	& 100-4000 nodes\\
$\D$			&  Ground coverage diameter & \\ 
			&of an FSO transceiver on HAP			& 15 km \\
$\W$			& Number of wavelengths in WDM & \\
			& technique				& 128 \\
$r^{max}$		& Data rate of a wavelength & 1 Gbps \\
$d_c$		& Maximum cloud size  & 10 km\\
$\C$			& Payload capacity of a HAP  & \\ 
			& in terms of FSO devices &10 devices\\
$\LHH$		& Maximal distance between HAPs			& 60 km\\
$\delta$		& BER threshold	& $10^{-3}$\\
\hline
\end{tabular}
\caption{Parameters of the dimensioning problem and their simulation values.}
\label{tab:param}
\end{table}

\section{Proposed dimensioning solution}
\label{sec:topo}

It is difficult to find an optimal HAP topology with the minimum total cost of HAPs and FSO transceivers in a single step. Therefore, we divide the design problem into three steps: i) \textit{minimum clustering}: identifying the minimum set of HAPs for serving all ground FSO nodes, ii) \textit{backup matching}: identifying a \bHAP for each \oHAP, iii) \textit{topology design}: linking HAPs together with the minimum number of FSO links in such a way that both the traffic demands of matrix \M~and their corresponding backup traffic are accommodated.

\subsection{Minimum clustering}
\label{sec:regroup-FSO}

To identify a minimum set of HAPs for serving all ground nodes, the ground nodes are organized into clusters, each of which is assigned a \sHAP. 
There are several constraints to be considered:
\begin{itemize}
	\item \textit{Limited coverage of FSO transceivers on HAPs}: ground nodes in a cluster should be inside the coverage diameter $\D$ of its \sFSO. Larger $\D$ allows a \sFSO~communicating with more ground nodes leading to fewer HAPs, but requires larger \sFSO~apertures thus higher transmit power to maintain the link quality. As a result, HAPs need to recharge more often. The optimal diameter $\D$ should be considered in taking count the trade off between HAP recharging cost and HAP cost. We reserve this optimization problem for a future work as fix $\D$ as a parameter in this research. 
	\item \textit{Limited number of wavelengths}: The number of ground FSO nodes per cluster is limited by $\W$, the number of wavelengths available according to the WDM technique.  	
\end{itemize}

\begin{figure}[tbh] \centerline{
\includegraphics[width=0.42\textwidth]{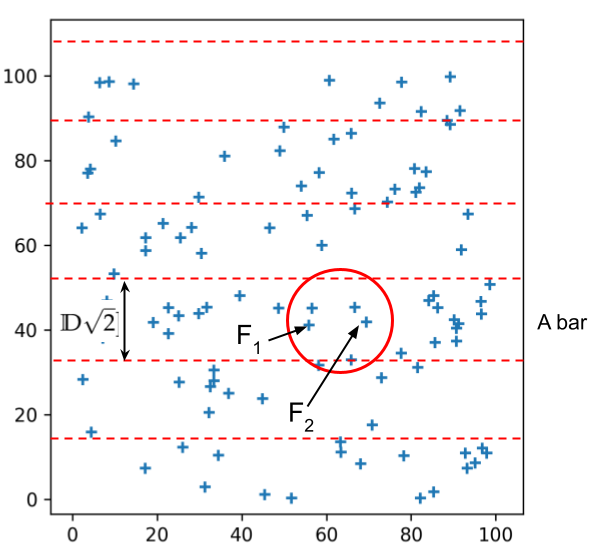}}
\caption{Clustering process with the ground region divided into bars.}
\label{fig:clustering}
\end{figure}

Let us describe the location of a ground FSO node by two coordinates $x$ and $y$ in a 2-dimension Cartesian coordinate system. Assume that all ground FSO nodes are distributed in the quarter where the $x$ and $y$ coordinates are all positive. We propose the following clustering algorithm:
\begin{itemize}
	\item Step 1: Divide the ground FSO region into non-overlapping bars parallel to the x-axis such that bar indexed $k$ contains the FSO devices with coordinates $\forall y \in [k \D \sqrt{2} \ldots (k+1)\D \sqrt{2}]$. See Figure \ref{fig:clustering}.      
	\item Step 2: We consider bars consecutively, from the index $k=0$ onward. For each bar, the FSO nodes are screened from the smallest to the largest $x$ coordinates: 
		\begin{itemize}
			\item Step 2.1: Find the FSO node, called $F_1$, with the smallest coordinate $x$ that has not been clustered.
			\item Step 2.2: Find the FSO node, called $F_2$, with the largest coordinate $x$ that has not been clustered such that: i) its distance to $F_1$ does not exceed $\D \sqrt{2}$, and ii) the number of ground FSO nodes located between $F_2$ and $F_1$ in the bar does not exceed $\W$. 
			\item Step 2.3: Create a new cluster including all FSO nodes in the bar between $F_1$ and $F_2$ inclusively. The \sHAP~for this cluster is placed at the coordinates $x=(x_{F_1} + x_{F_2})/2$ and $y=(k+1/2) \D \sqrt{2} $.
			\item Step 2.4: If the cluster is not full, i.e., the number of FSO nodes is still under $\W$, then put into the cluster the FSO nodes that are in the coverage diameter $\D$ of the newly identified \sHAP,~even if those nodes do not belong to the current bar. The process is repeated until the cluster is full. 
			\item Step 2.5: If there are still FSO nodes unclustered in the bar, repeat step 2.1 to create other clusters.
		\end{itemize}
\end{itemize}

\begin{figure}[tbh]\centerline{
\includegraphics[width= 0.4\textwidth]{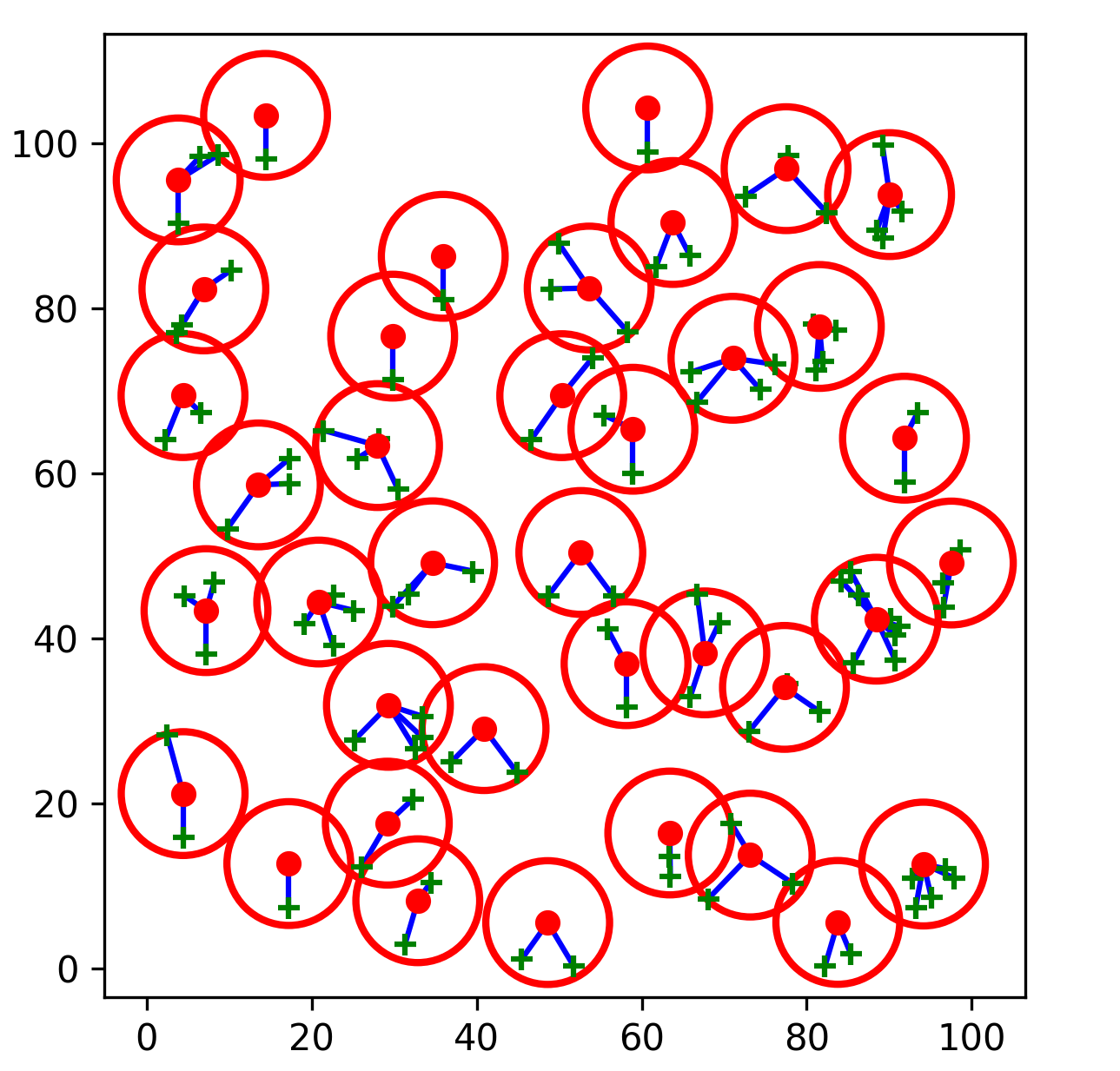}}
\caption{Example of the clustering result.}
\label{fig:cluster-example}
\end{figure}

The results of the algorithm are the projected locations of all \sHAP s on the ground and the set of ground FSO nodes served by each HAP. Figure \ref{fig:cluster-example} shows the clustering result for the set of ground FSO nodes shown in Figure \ref{fig:clustering}. Each cluster is represented by a circle in which the centroid is the projected location of its \sHAP. Ground FSO nodes are represented by green + symbols in the figure. We can remark that more clusters are formed where the density of the ground FSO nodes is high, while fewer or even no clusters are resulted where the density of  ground FSO nodes is low.

\subsection{Backup matching}
In this step, we find a backup HAP for each \sHAP. Assume the maximum cloud size is $d_c$, the \bHAP~should be at a distance of at least $2 \times d_c$ from the primary one.

Our idea is that, starting from the set of HAPs found from the clustering step, we match them in pairs so that HAPs in a pair backup each other. The matching problem can be solved using Edmonds' Blossom algorithm \cite{edmonds_1965} which finds a maximum cardinality matching in a graph. Applying to our problem, the graph contains all HAPs as vertices and edges are inter-HAP links whose lengths are greater than $2 \times d_c$ but smaller than $\LHH$, the maximum allowable length for an inter-HAP link. The maximum matching provides the largest number of pairs of primary and backup HAPs. If the algorithm fails to find a matching for  a HAP, we add a new HAP as the backup HAP for it. The newly added backup HAP must be at $2 \times d_c$ distance from the primary HAP.

After matching backup, a number of wavelengths need to be reserved on the link between a \oHAP~and its \bHAP~for carrying backup upstream and downstream traffic. In Figure \ref{fig:backupsol}, when $H_B$ backs up $H_A$, the number of wavelengths to be reserved on link $H_B \rightarrow H_A$~is equivalent to the number of ground FSO nodes served by $H_A$ because the upstream traffic from a ground FSO node travels on a single wavelength over link $H_B \rightarrow H_A$. Similarly, the same number of wavelengths must be reserved on the opposite link $H_A \rightarrow H_B$ for downstream traffic. 

Since $H_A$ also backs up $H_B$, a number additional wavelengths equivalent to the number of ground FSO nodes served by $H_B$ should be reserved on the links $H_A \rightarrow H_B$ and $H_B \rightarrow H_A$. As a result, the link $(H_A, H_B)$ can accommodate backup traffic for both $H_A$ and $H_B$ only if the total number of ground FSO nodes of $H_A$ and $H_B$ does not exceed $\W$.

\subsection{Topology design and wavelength routing}
\label{sec:topoHAP}

Once the locations of primary HAPs are identified and backup HAPs have been assigned to them, the remaining issue is to determine the inter-HAP links to be deployed and the lightpaths  to be setup between HAPs in order to accommodate all traffic requests. Since the wavelengths dedicated to backup traffic are reserved during the backup matching step, the remaining traffic to be accommodated is the primary traffic. The amount of primary traffic to be sent between two HAPs is the sum of the demanded bandwidth between their ground FSO nodes. Therefore, the number of lightpaths to be setup between a pair of HAPs is  the ratio of the demanded bandwidth between the HAPs and the bit rate $r^{max}$ of a wavelength. The topology design problem is stated as follows: \\
Given:
\begin{itemize}
	\item Locations of HAPs obtained from the clustering step and those added in the backup matching step.
	\item \MHAP: set of traffic demands between pairs of HAPs. Each demand is represented by a tuple $(s, d,n)$, where $s$, $d$ and $n$ are, respectively, the source HAP, destination HAP and number of lightpaths needed to carry traffic from $s$~to $d$. 
	\item Some parameters as given in Table \ref{tab:param}. 
\end{itemize}
Constraints:
\begin{itemize}
	\item All end-to-end lightpaths should have BER under $\delta$.
	\item Wavelength continuity constraint along each lightpath.
\end{itemize}
Output:
\begin{itemize}
	\item We need to identify the necessary inter-HAP links and lightpaths for carrying all traffic in \MHAP~such that the total number of FSO transceivers to be carried on HAPs is minimized. 
\end{itemize}

The idea of the proposed algorithm is that: from a set of possible inter-HAP links, we find routes for demands in \MHAP one by one, and incorporate the links on the routes in the topology. The links already included in the topology are assigned small weights so that they will be prioritized in subsequent demands, thus new links are involved only when there is no possible route within the current topology. 

Wavelengths are assigned to routes based on the \textit{least used} strategy. The least used wavelength is defined as the wavelength that carries traffic on the smallest number of inter-HAP links.

Below are some notations used in the algorithm:

\begin{itemize}

\item $G_w$ denotes the directed graph, where vertices represent HAPs, and arcs represent all possible directional inter-HAP links with wavelength $w$ available.
\item $nFreeW(e)$ denotes the number of free wavelengths of directional link $e$. 
\item $c_e$ denotes the weight assigned to directional link $e$,
	\begin{equation}
	c_e=\left \{
	\begin{array}{ll}
	100 			   & \mbox{if } e \mbox{ has never been used} \\
	1- \frac{nFreeW(e)}{\W} & \mbox{otherwise}\\
	\end{array}
	\right .
	\label{eq:weight}
	\end{equation}
\item $T$ denotes the topology of the HAP network during construction. Initially, $T$ contains all bidirectional inter-HAP links connecting \oHAP s and their \bHAP s. On these links, wavelengths for backup purpose were marked as used.
\item $C(u)$ denotes the number of FSO devices to be on HAP $u$ according to topology $T$. $C(u)$ includes all serving FSO devices of HAP $u$ and one FSO device as the endpoint for each bidirectional inter-HAP link adjacent to $u$.
\end{itemize}

\begin{algorithm}
\caption{Topology design and wavelength routing algorithm}
\begin{algorithmic}[1]
\Function{Main}{}
\ForAll{ $r(s,d,n) \in$ \MHAP }
	\State \Call{Routing-one-demand}{s,d,n}
\EndFor
\EndFunction

\Function{Routing-one-demand}{s, d,n}
\State $T \leftarrow \emptyset$ \;
 \State LP-set $\leftarrow$ NULL \; \Comment{set of lightpaths to be found}
 \State nb-trial-w $\leftarrow$ 0 \; \Comment{number of trial wavelengths}
\Repeat
	\State $w \leftarrow$ the least used wavelength \; 
	\State nb-trial-w ++
	\Repeat
	\State $p \leftarrow$ \Call{SP-Constraint}{$G_w, s, d$} \;
	\State valid-path $\leftarrow$ true \;
	\ForAll{$(u,v) \in p$ \AAnd $(u,v) \notin T$} 
		\If{$C(u) \geq \C$ \AOr $C(v) \geq \C$}
			\State valid-path $\leftarrow$ false
			\State $G_w \leftarrow G_w \setminus (u,v)$
			\State break;
		\EndIf
	\EndFor
	\Until{ $p$ not found \AOr valid-path = true} 
	\If{$p$ not found}  next \; \Comment{try the next wavelength}
	 \EndIf
	\If{valid-path} \Comment{Assign wavelengths}
		\ForAll{ $(u,v) \in p$} 
			\State $T \leftarrow T \cup (u,v) $ \Comment{add new arcs to $T$}
		\EndFor
		\State LP-set $\cup$ path $p$ wavelength $w$.
		\State remain-LP $\leftarrow n-1$
		\While{ remain-LP$>0$ \AAnd nb-trial-w$ <\W$}
			\State $w_k \leftarrow$ the next least used wavelength
			\If{$w_k$ is available along $p$}
				\State LP-set $\cup$ path $p$ wavelength $w_k$
				\State remain-LP $\leftarrow$ remain-LP -1
			\EndIf
		\EndWhile
		\If{remain-LP $>0$} 
			\State \MHAP $\leftarrow$ \MHAP $\cup$ ($s$,$d$, remain-LP)
		\Else 
			\State break
		\EndIf
	\EndIf
	
\Until{nb-trial-w$=\W$}
\If{remain-LP$ >0$} 
	\State  \Return false  \Comment{Reject demand} 
\EndIf
\State \Return LP-set \;
\EndFunction

\end{algorithmic}
\label{alg:routing}
\end{algorithm}

Algorithm \ref{alg:routing} shows the pseudo-code of the proposed algorithm. For each demand $(s,d,n) \in $ \MHAP, we find $n$ lightpaths from $s$ to $d$, each one uses a single wavelength and possibly take a individual route. The main steps are explained bellow: 
\begin{itemize}
	\item Step 1 (lines 13--23 of Algorithm \ref{alg:routing}): Assume  that the current \emph{least used} wavelength is $w$, we find the weighted shortest path from $s$ to $d$  in $G_w$ using Algorithm \ref{alg:djk-constraint}. Algorithm \ref{alg:djk-constraint}, proposed in \cite{TruongNICS2019}, is a modified Dijkstra algorithm, which finds the shortest path having end-to-end BER under threshold $\delta$. The found path is called $p$.  Each arc of $p$, that is not in $T$ yet, is checked to see if both HAPs adjacent to it having free payload capacity for installing an additional FSO device for the arc. If at least one HAP violates the constraint, the arc is removed from $G_w$, and another path will be sought using Algorithm \ref{alg:djk-constraint} again. This step could be repeated until no violation is found, or no path is found. 
	\item Step 2: If no path is found on $G_w$ for the current demand, then try to find a path on $G_j$, where $j$ is \textit{the next least used wavelength} and so on (line 24). When all wavelengths have been screened but no path is found, this demand is rejected (lines 46--47).
	\item Step 3: When no violation is found, the path is good for the current demand, wavelengths are assigned to it to make lightpaths:					\begin{itemize}
			\item The new arcs of $p$ are added to $T$ and the first lightpath for the demand $(s,d,n)$ is recorded as using wavelength $w$ along path $p$ (lines 26--30). 
			\item If the number of requested lightpaths $n>1$, the remaining number of lightpaths to be created is $remain-LP = n-1$. We try to make the remaining lightpaths by using always path $p$ but other wavelengths starting from the \emph{least used} wavelength available first (lines 31--38).
			\item If there are still not enough available wavelengths to route the entire demand over $p$, the un-routed lightpaths should follow a different path. We consider the remaining part of the current demand as a new one with parameters $(s, d, remain-LP)$ and add it to \MHAP for later handling (lines 39--40).
		\end{itemize}
\end{itemize}

At the end of the algorithm, we obtain the topology $T$ for the HAP network and all lightpaths for the accepted demands in \MHAP.


\begin{algorithm}
\caption{Shortest path with BER constraint}
\begin{algorithmic}[1]
\Function{SP-Constraint}{\G, s, d} 
\State Q $\leftarrow$ vertex set of \G\;
\ForAll{vertex $v \in Q$ }
	 \State dist[v] $\leftarrow$ INFINITY\;
	\State prev[v] $\leftarrow$ UNDEFINED\;
\EndFor
	\State dist[s] $\leftarrow$  0\;
\While{$Q \neq \emptyset$}
	\State  u $\leftarrow$ vertex in $Q$ with min dist[u]\;  
         \ForAll{neighbor $v$ of $u$}           
	       \State alt $\leftarrow$ dist[u] + length(u, v)\;
                \If {alt $<$ dist[v]  \AAnd BER-e2e(s,u,v)}               
                	    \State   dist[v] $\leftarrow$ alt\;
    	              \State prev[v] $\leftarrow$ u\;
	         \EndIf
    	\EndFor
\EndWhile
\Return dist[d], prev[d] \;
\EndFunction

\Function{BER-e2e}{s,u,v}
	\State $prod \leftarrow (1-\BER{(u,v)})$\;
        \For{$\ell \in$ current path from $s$ to $v$}
            \State $ prod \leftarrow prod \times (1-\BER{\ell})$\;
	\EndFor
        \If{ $prod > 1-\delta$} 
	     \State \Return true\;
       \Else
	     \State \Return false\;       
	\EndIf
\EndFunction
\end{algorithmic}
\label{alg:djk-constraint}
\end{algorithm}

\begin{figure} [htb]
\includegraphics[width=0.5 \textwidth]{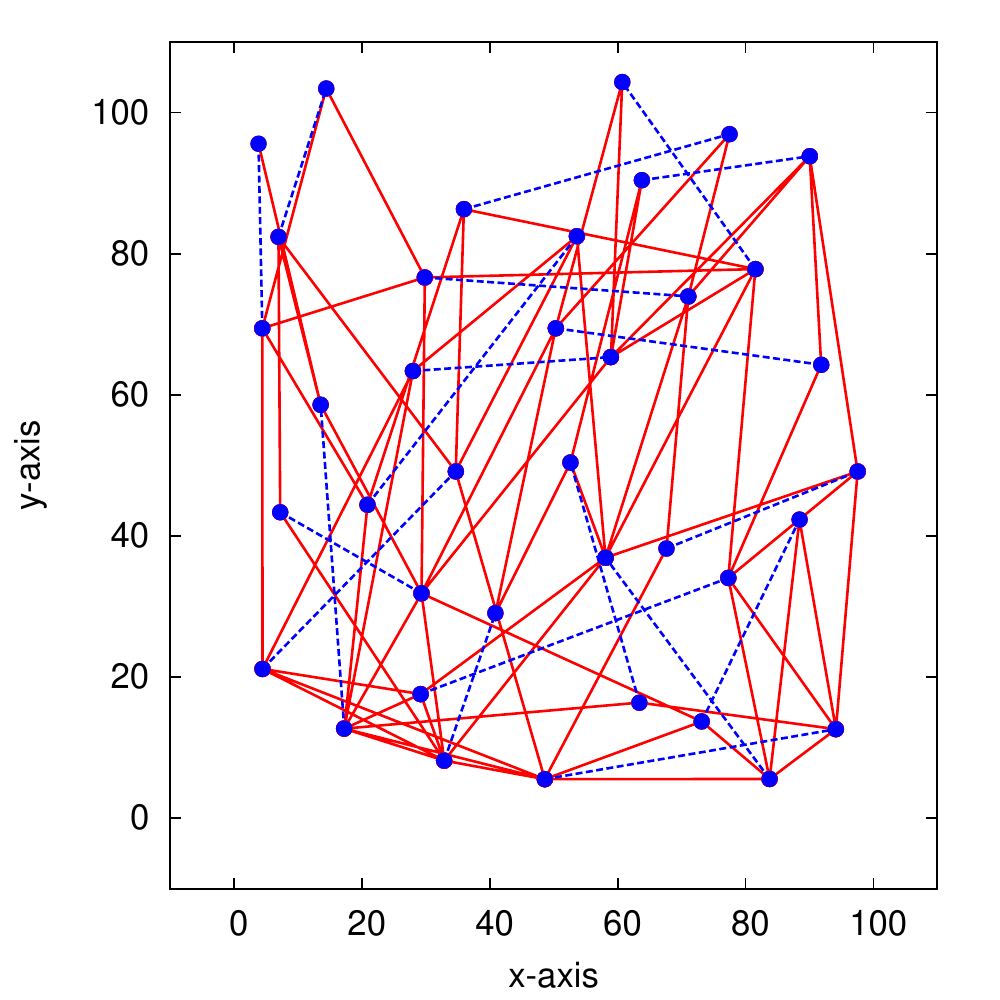}
\caption{HAP network designed by the proposed algorithms. Blue dots are HAPs acting as backup HAPs, blue dashed lines are inter-HAP links between a pair of primary and backup HAPs. }
\label{fig:net-example}
\end{figure}

Figure \ref{fig:net-example} shows the topology of the HAP network designed by the proposed algorithm for the test case shown in Figure \ref{fig:cluster-example}. Blue dots are HAPs providing backup function, and blue dashed lines are inter-HAP links between a pair of primary and backup HAPs. 

\section{Numerical results}
\label{sec:simulation}

We implemented the algorithms in Section \ref{sec:topo}, including FSO clustering, backup matching and topology design steps. Since there is no previous work on dimensioning problem for survivable FSO networks using HAPs against optical uplink and downlink failures, thus we can not evaluate our dimensioning solution against any previous one. We will thus concentrate on analyzing  the impact of protection on topologies by comparing the topologies with protection against the topologies without protection. Topologies without protection were obtained by running only the clustering and topology design steps.

The comparisons were made on different test cases. The ground FSO node locations were generated randomly on a square surface of 100km $\times$ 100km, which is the size of a metropolitan area. The total number of ground FSO nodes was generated randomly in the range of 100 to 4000 nodes.  The following parameters were set based on the technological and theoretical review in \cite{Optical-HAP-2010} on optical communications for HAP. The transmission rate per wavelength $r^{max}$ was set to 1~Gbps. The aperture of \sFSO~devices was set approximately $40^{\circ}$; thus, the ground coverage diameter of a \sFSO~was $\D \sim 15 $ km. Traffic requirement was generated randomly between ground FSO nodes, but the total incoming or outgoing traffic of a ground FSO node does not exceed 1 Gbps, which is the capacity of a single wavelength. 

The BER of an inter-HAP link was calculated according to air turbulence model using Gamma-Gamma distribution in \cite{ngoc2019} with moderate turbulence condition. BER threshold for an end-to-end lightpath as well as for a single inter-HAP link was set as  $\delta = 10^{-3}$, so that the errors could be recover by current FEC techniques. According to the BER computation model, inter-HAP link longer than 60 km having BER greater than $\delta=10^{-3}$, therefore, $\LHH$ was set to 60 km. The wavelength density of the WDM technique was $\W=128$ wavelengths. The recapitulation of the parameter values is shown in Table \ref{tab:param}.

In all test cases, we observed no demand rejection either with or without protection. We will discuss the impact of protection on the number of HAPs and the number of FSO devices on HAPs to be invested. We also analyze the resource occupation with and without protection.
\subsection{Number of additional HAPs due to protection}

\begin{figure}[tbh]
\includegraphics[width=0.5 \textwidth]{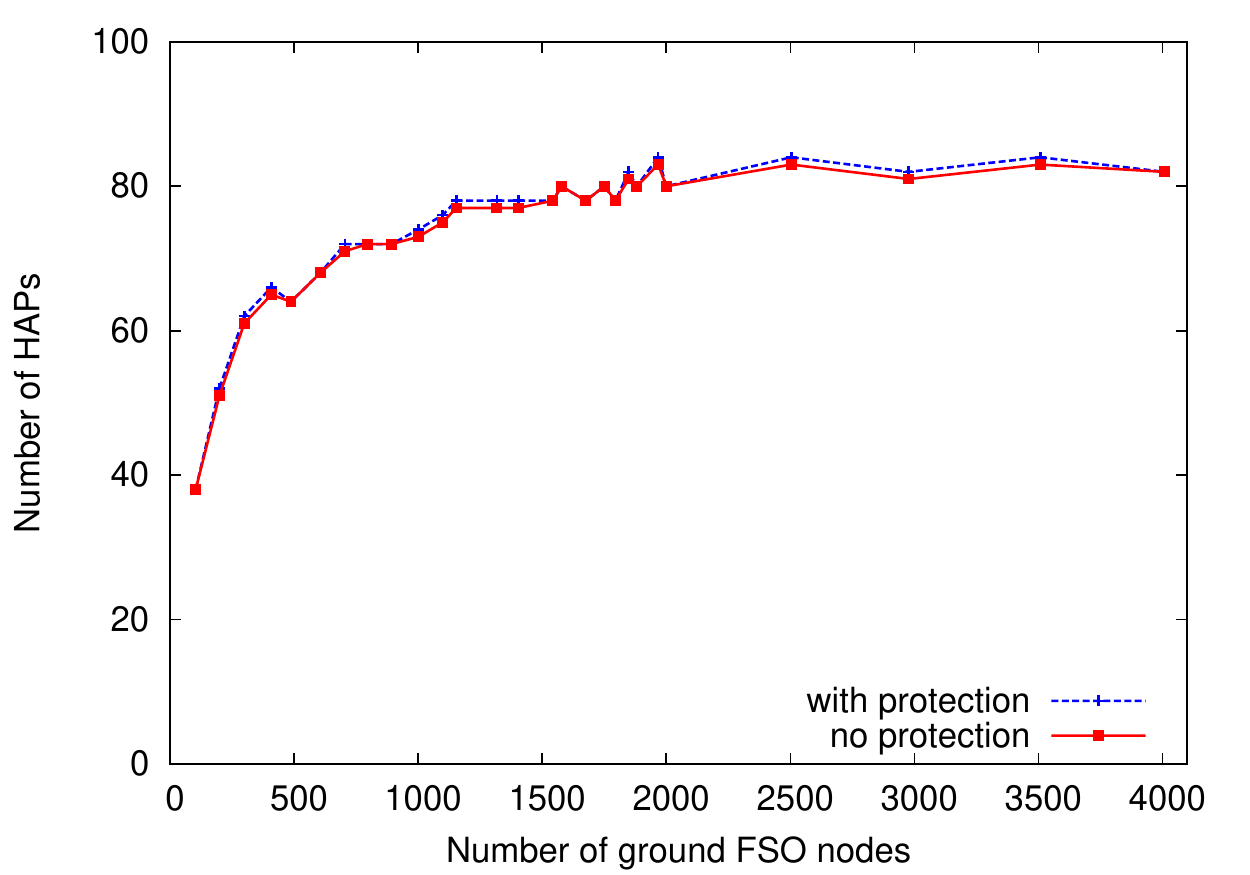}
\caption{Number of HAPs}
\label{fig:nb-HAP}
\end{figure}

Figure \ref{fig:nb-HAP} shows the number of HAPs as a function of the number of ground nodes. This shows that protection requires \textit{at most one more HAP} in all test cases. The reason is that: if HAPs still have free payload for installing backup serving FSO devices, \sHAP s of different clusters backup each other. The additional HAP is only involved when the number of clusters is impaired. Thus, \emph{the protection does not lead to a higher cost of HAPs if the HAPs are not too loaded prior protection.}

\subsection{Number of FSO devices on HAPs}
\begin{figure}[tbh]
\includegraphics[width=0.5 \textwidth]{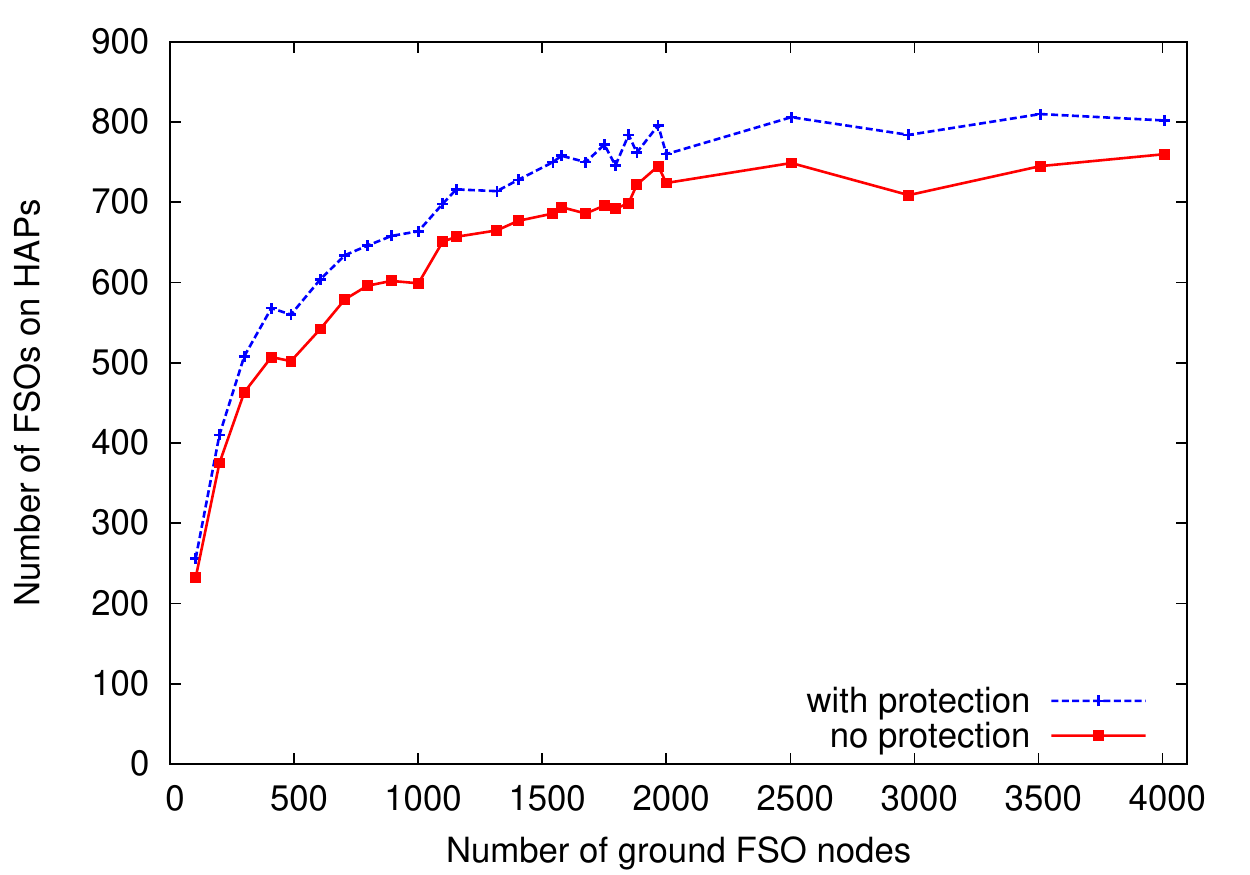}
\caption{Total number of FSO devices carrying by all HAPs}
\label{fig:fso-HAP}
\end{figure}

Figure \ref{fig:fso-HAP} shows the total number of FSO devices to be installed on HAPs with and without protection. Evidently, cases with protection require additional FSO devices to serve ground nodes in the protected zones and to connect the primary and backup HAPs. Nonetheless, the total number of FSO devices to be deployed on HAPs increases only by 5\%--12\% owing to protection. 

\begin{figure}[tbh]
\includegraphics[width=0.5 \textwidth]{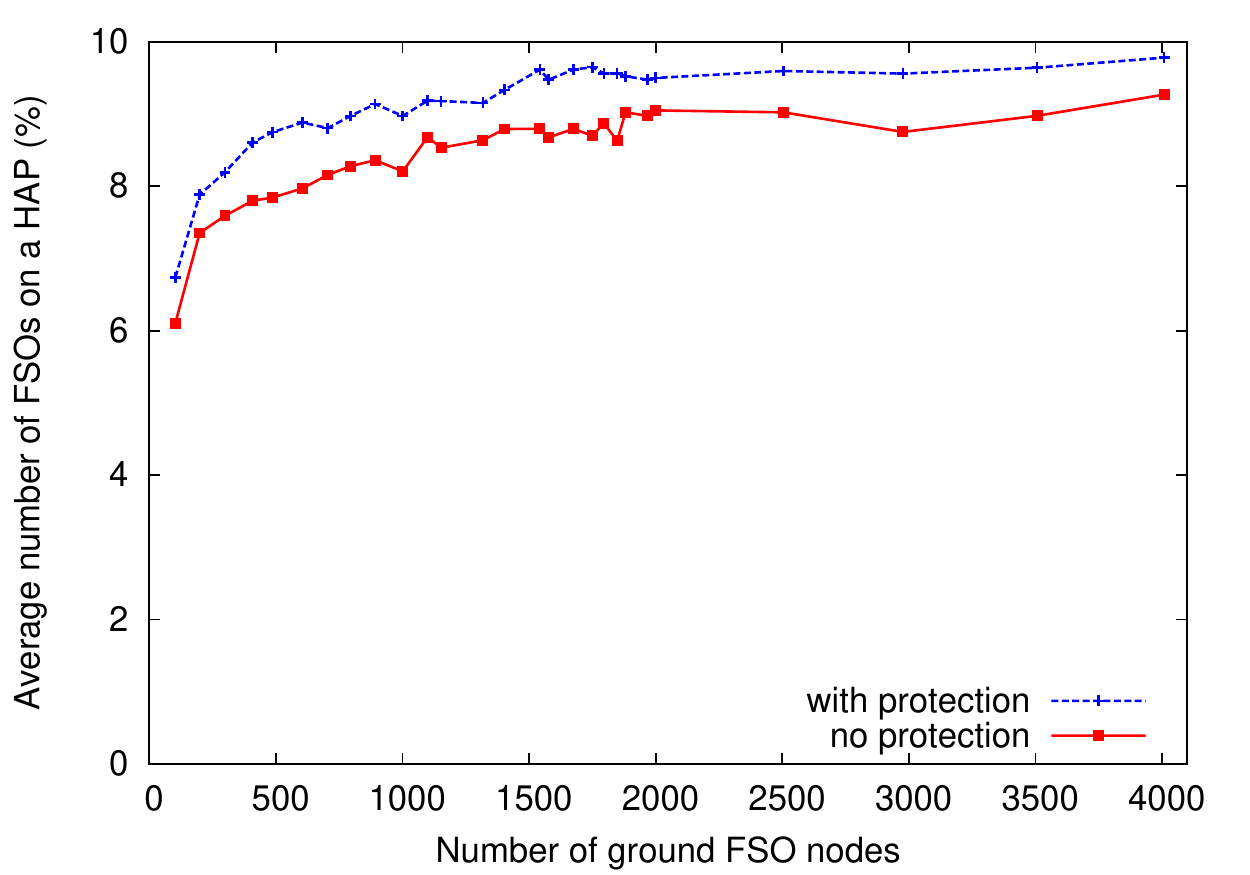}
\caption{Average number of FSO devices carrying by one HAP}
\label{fig:mean-FSO}
\end{figure}

Figure \ref{fig:mean-FSO} shows the average number of FSO devices that a HAP carries with and without protection. We observe that, on average, each HAP has to carry from 0.45--0.95 more FSO devices for  protection purpose. 

\subsection{Number of additional inter-HAP links and occupancy due to protection}

\begin{figure}[tbh]
\includegraphics[width=0.5 \textwidth]{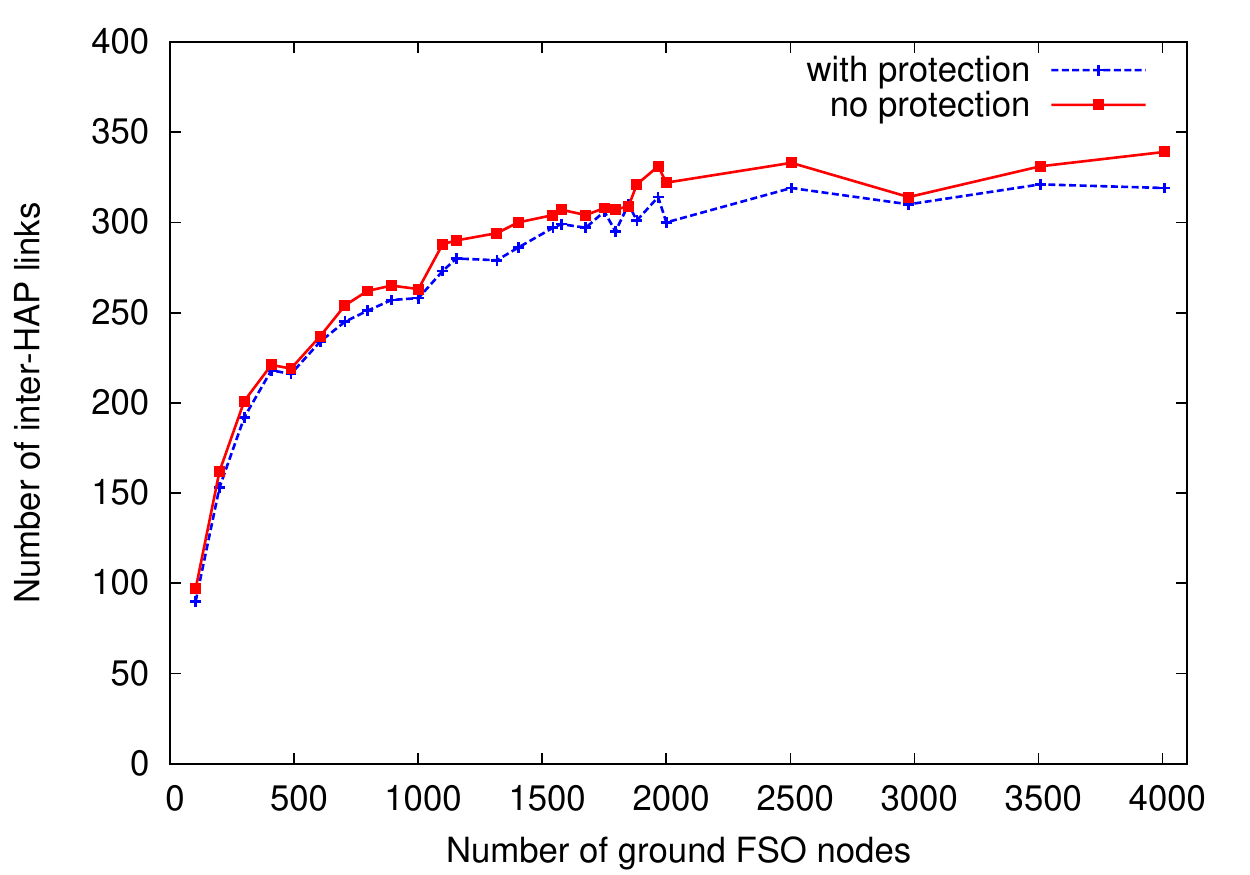}
\caption{Number of inter-HAP links}
\label{fig:inter-HAP-links}
\end{figure}

\begin{figure}[tbh]
\includegraphics[width=0.5 \textwidth]{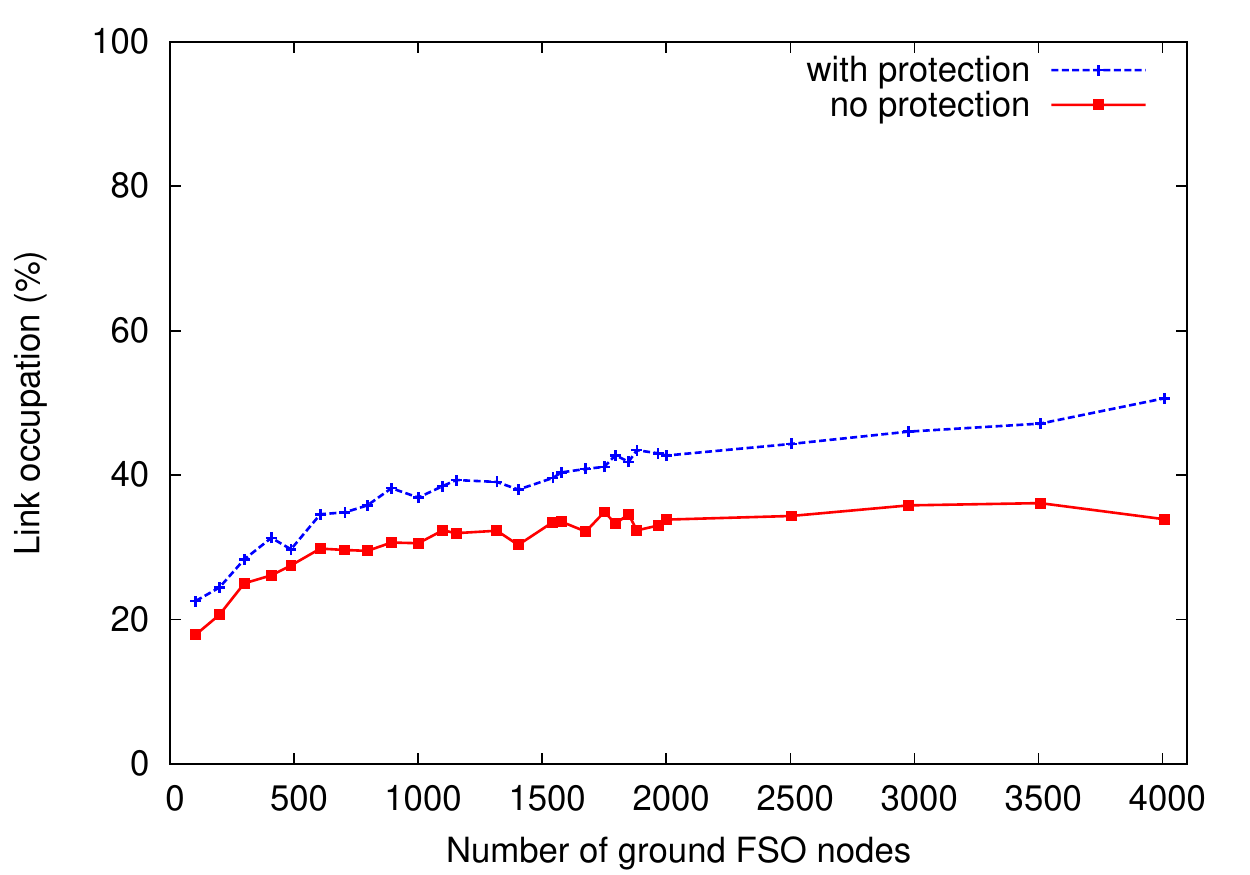}
\caption{Link occupation}
\label{fig:link-usage}
\end{figure}

Figure \ref{fig:inter-HAP-links} shows the number of inter-HAP links when the number of ground nodes increases. Since traffic demands were generated between all ground nodes, the volume of traffic increases proportionally with the number of ground nodes. As a result, more inter-HAP links are involved in networks with larger number of ground nodes.

Counter intuition, the statistics demonstrate that the topology with protection generally uses fewer links than the topology without protection. The difference becomes clearer when the number of ground nodes increases. We believe that this phenomenon is due to the characteristics of the proposed algorithm. In the backup matching step, a number of inter-HAP links connecting pairs of primary and backup HAPs are included in the initial topology $T$. Consequently, the topology design algorithm frequently uses these links,  resulting in fewer added links than in the no-protection scenario. We believe that by modifying the topology design algorithm, it is possible to achieve a topology without protection with fewer links.

The analysis of link occupation confirms that links are exploited more intensively in topologies with protection. We define link occupation as the ratio between the number of wavelengths used and the total number of wavelengths of the entire network:
\begin{equation}
\mbox{link occupancy}=\frac{\mbox{Number of  wavelengths used on all links}}{\mbox{Number of inter-HAP links} \times 2 \W.}
\end{equation}
Coefficient 2 in the denominator is introduced because a link is bidirectional, and wavelengths are counted separately in opposite directions. 

Figure \ref{fig:link-usage} shows that link occupation in topology with protection is clearly higher than link occupation in topology without protection mostly when the number of ground FSO nodes increases.

\subsection{Wavelength cost for protection purpose}

\begin{figure}[tbh]
\includegraphics[width=0.5 \textwidth]{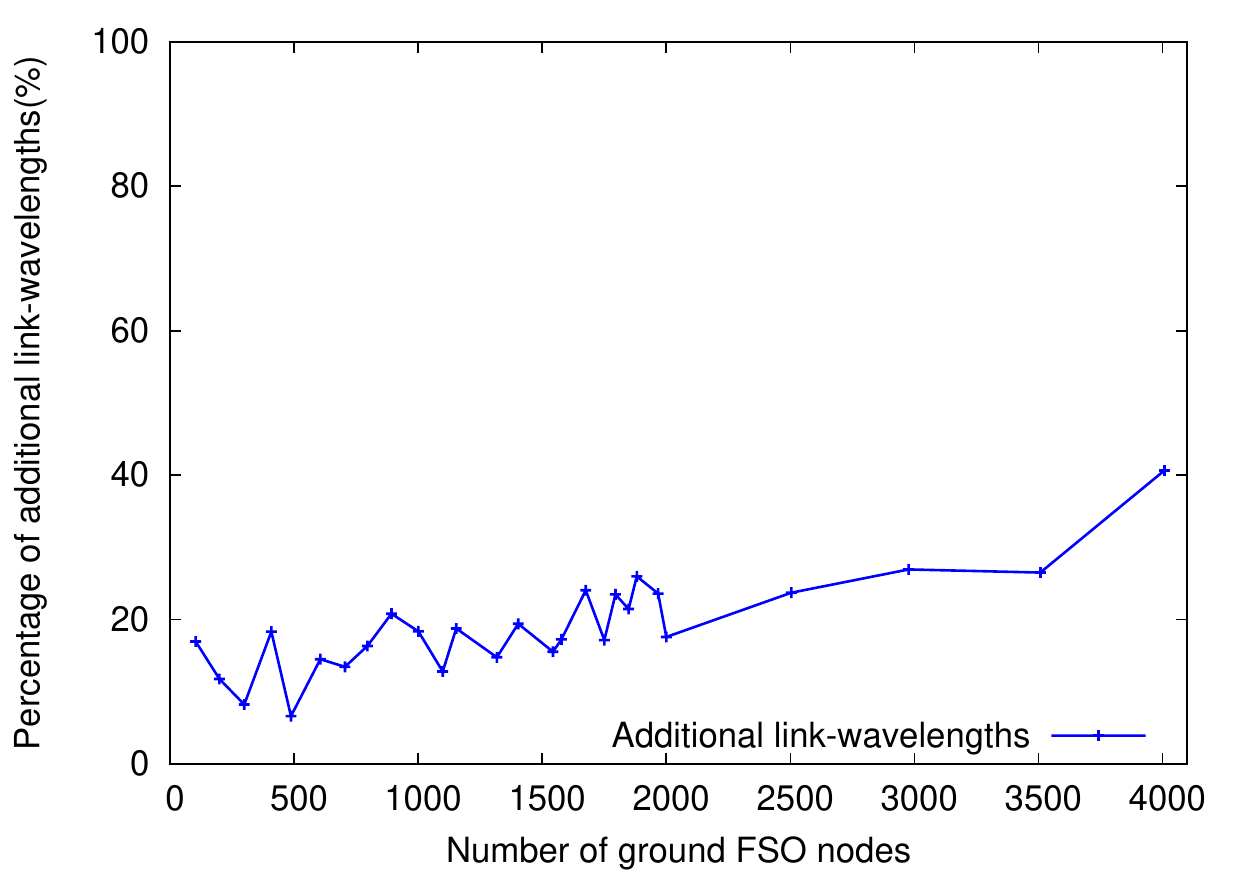}
\caption{Percentage of additional wavelength resources for protection in comparison with no protection}
\label{fig:ressource}
\end{figure}

We evaluated the impact of protection on the quantity of wavelength resources in the network topology.  Let \emph{link-wavelength} be a wavelength of a link. The number of link-wavelengths increases when more wavelengths are used in many links. We calculated the total number of link-wavelengths used in all inter-HAP links of the topologies with and without protection for each test case.  Figure \ref{fig:ressource} shows the percentage of additional link-wavelengths used for protection. The topologies with protection use from 8\% to 40\% more link-wavelengths than those without protection. The amount of additional link-wavelengths is modest because in the case with protection, demands travel over only one more link (between the backup and primary HAPs) compared to the case without protection. 

\section{Conclusions}
\label{sec:conclusion}
HAPs as intermediate systems for long-distance FSO communication are promising since they provide broadband communication. The presence of clouds in the middle of the uplinks and downlinks between HAPs and the ground is a major obstacle. In this research, we proposed a switching and protection mechanisms as well as a network dimensioning solution for survivable end-to-end communication against clouds. Data travel on both primary and backup paths using uniquely FSO communication technology. The simulation results show that a network with survivable capability is not necessary much more expensive than that without protection in terms of HAP and FSO investment costs, meanwhile connection availability can reach to 97.75\% based to Europe cloud coverage data. 

The network dimensioning would be better optimized if the apertures of different \sFSO s could be adjusted independently leading to various cluster sizes. In a future work, we will address the \sFSO~apertures optimization problem.

\section*{Acknowledgements}
This research was funded by the Vietnam National Foundation for Science and Technology Development (NAFOSTED) under grant number 102.02-2018.305. 

\bibliographystyle{elsarticle-num}
\bibliography{FSO-HAP}

 \end{document}